# Towards an intelligence based conceptual framework for e-maintenance

Abdessamad Mouzoune

QSM Laboratory - CEDOC / EMI,
Mohammed V University – Agdal, Rabat, Morocco

Mouzoune_abdessamad@yahoo.fr

**Abstract.** Since the time when concept of e-maintenance was introduced, most of the works insisted on the relevance of the underlying Information and Communication Technologies infrastructure. Through a review of current e-maintenance conceptual approaches and realizations, this paper aims to reconsider the predominance of ICT within e-maintenance projects and literature. The review brings to light the importance of intelligence as a fundamental dimension of e-maintenance that is to be led in a holistic predefined manner rather than isolated efforts within ICT driven approaches. As a contribution towards an intelligence based e-maintenance conceptual framework, a proposal is outlined in this paper to model e-maintenance system as an intelligent system. The proposed frame is based on CogAff architecture for intelligent agents. The architecture decomposes the agent into three layers (perception, central processing and action) with a hierarchical vision of three levels. Within the proposed frame, more importance was reserved to the environment that the system is to be continuously aware of: Plant Environment, Internal and External Enterprise Environment and Human Environment. In addition to the abilities required for internal coherent behavior of the system, requirements for maintenance activities support are also mapped within the same frame according to corresponding levels (strategic, tactical and operational management). A case study was detailed in this paper sustaining the applicability of the proposal in relation to the classification of existing e-maintenance platforms. The case study also underlined two issues of existing e-maintenance solutions regarding real integration of the system into human teams and more integration at the management strategic level. However, more work is needed to enhance exhaustiveness of the frame to serve as a comparison tool of existing e-maintenance systems. At the conceptual level, our future work is to use the proposed frame in an e-maintenance project that will have the two mentioned issues as an objective.

# 1.    Introduction

E-maintenance has been discussed in many maintenance related literature with different perspectives. E-maintenance basically refers to the integration of the information and communication technologies (ICT) within the maintenance strategy or plan (Chowdhury and Akram 2011).

While perpetual development in ICT is providing new opportunities to the existing cultures of maintenance, e-maintenance seems to be predominately ICT-driven, meaning that the creation of innovative engineering solutions advanced the practice and research of e-maintenance (Haftor et al.2010).

Reconsidering the distributed artificial intelligence environment generating e-maintenance system, this work aims firstly to highlight the importance of intelligence as a fundamental dimension that conceptual frameworks for e-maintenance may gain by being based on.

The rest of the paper is organized as follows: Section2 reviews e-maintenance as an ICT based system. Section 3 explores e-maintenance as an intelligent system and a frame is then proposed to map it from an intelligence point of view. In section 4 a case study is detailed using the proposed frame. In the last sections, our proposal is discussed defining future work to be done.

# 2.    Background: ICT driven e-maintenance models

Maintenance is defined by the European Committee for Standardization (EN 13306:2001) as the combination of all technical, administrative and managerial actions during the life cycle of an item intended to retain it in, or restore it to, a state in which it can perform the required function (function or a combination of functions of an item which are considered necessary to provide a given service).

The same standards defines maintenance management as all the activities of the management that determine the maintenance objectives or priorities, strategies, and responsibilities and implement them by means such as maintenance planning, maintenance control and supervision, and several improving methods including economical aspects in the organization.

Maintenance Management includes corrective, preventive and proactive maintenance, inventory and procurement, work order system, computerized maintenance management systems (CMMS), reliability centred maintenance, total productive maintenance, financial optimization, technical training, and continuous improvement.

From an organizational point of view, maintenance management must align actions at three levels of business Activities. Referring to (Márquez, Adolfo Crespo. 2007)  ), this means:

The Strategic level establishes maintenance priorities in accordance with business priorities.   This priorities transformation materialized by a generic maintenance plan will establish critical targets in current operations. In addition, maintenance management at this level is responsible to decide on skills and technologies requirements to improve maintenance effectiveness and efficiency.

The tactical level of maintenance management is responsible of the correct assignment of maintenance resources  to fulfill the maintenance plan. Hence, detailed maintenance requirements planning and scheduling are established at this level. Tactical maintenance policies are to be improved with experience.

The operational level ensures that the maintenance tasks are carried out by skilled technicians, in the time scheduled, following the correct procedures, and using the proper tools. This level of maintenance management is also responsible of data to be recorded for diagnosis and/or prognosis purposes.

In addition, Information and Communication Technologies (ICT) is considered as one of the maintenance pillars beside maintenance engineering methods and organizational ones (Crespo Marquez, Adolfo, and Jatinder N.D. Gupta.2006). In fact, CMMS are among the first historical steps of maintenance information systems (Rasovska et al. 2007). With the increasing development of ICT, other steps followed namely tele-maintenance, e-maintenance and semantic e-maintenance (s-maintenance). The main issue of these steps globally referred to as "e-maintenance" is integrating different systems relative to maintenance into one information system.

Since the time when concept of e-maintenance was introduced, most of the works insisted on the relevance of the underlying ITC infrastructure. Referring to (Arnaiz et al. 2010), technology sources are basically :

Devices miniaturization increasing the ways data can be acquired including mobile systems.

Extension of communication technologies including wireless and usage of the Internet as a main distributed platform for business operation.

Supported by specific information standards for systems interoperability, such as MIMOSA and OSA-CBM, ICT, technologies include:

- New Sensor Systems such as smart sensors—MEMS, performing Condition-Based e-maintenance and RFID (Radio Frequency Identification Device), storage of conventional data

- PDA (Personal Digital Assistant), SmartPhones, Graphic tablets, harden laptops, and other mobile devices,

- Wireless technologies mainly Bluetooth (IEEE 802.15.1) and ZigBee (IEEE 802.15.4) ,

In short, (Muller et al. 2008) attributes the e-maintenance emergence its self to two main factors that are (1) E-technologies allowing optimization of maintenance related workflow; and (2) Integrating business performance, and dealing with openness, integration, and collaboration with the other services of the e-enterprise.

Ample literature exists on design, development, and integration of e-maintenance within enterprise processes , including theories and practical applications, without any consistent definition as shown by a set of different e-maintenance definitions listed in (Arnaiz et al. 2010). In relation to our topic, none of those definitions explicitly mention intelligence as a fundamental characteristic of e-maintenance. Nevertheless, each implicitly recognizes such an importance especially when it comes to real time data processing and automation of some maintenance tasks. When explicit mention of intelligence is needed, an adjective is added to corresponding system as in "Intelligent Maintenance System" reported in an editorial of "Computers in Industry" (2006)): intelligence is defined as ''the ability to monitor plant floor assets, link the production and maintenance operations systems, collect feedbacks from remote customer sites, and integrate it in upper level enterprise applications.''

In fact, from tele-maintenance to semantic e-maintenance, the major concern was integration, communication and interoperability in an ICT driven e-maintenance development. ICT is a fundamental dimension that enhances cooperative and collaborative characteristics of maintenance activities. Being so, ICT are at the base of every definition of e-maintenance.

The predominance of ICT dimension in textual frameworks represented by different definitions of e-maintenance is replicated on conceptual ground. Nevertheless, this predominance is more and more rectified by some conceptual frameworks to include strategic vision, business processes and organizations as in (Iung et al. 2009): Following the Zachman framework founded as an information system architecture framework (Zachman, John A. 1999) and later presented as an enterprise architecture, a comprehensive conceptual e-Maintenance framework is proposed based on five abstraction levels. These are (1) E-maintenance strategic vision (business and goal) which supports the "scope", (2) E-maintenance business processes which support the Business view, (3) E-maintenance organization which supports the Architect's view, (4)E-maintenance service and data architecture which supports the designer's view, and(5) E-maintenance IT infrastructure supporting the builder's view. Another conceptual framework aiming same objective to limit such an ICT predominance is presented in (Haftor et al. 2010) based on Information Logistics as a Driver for Development. Information logistics are defined as the provision of information for the knowledge worker within the correct context, enabling him or her to make the right decision (Logistics - Wikipedia. 2012).

In general terms, though known frameworks, recognize the need for intelligence within e-maintenance systems, none of them goes beyond a shallow model of that dimension. Meanwhile, several industrial and academic e-maintenance platforms have been developed and are in use today. Within such platforms, artificial intelligence in general and machine learning algorithms in particular are by far important enablers to model maintenance tasks such as condition monitoring, prognostics, prognosis and decision support. For PROTEUS platform, a generic artificial intelligence (AI) template is defined in (Déchamp et al. 2004) to specify how AI tools must be integrated into the platform for a given diagnosis or prognosis task. More generally, (Muller et al. 2008) reports a state of the art of contributions in the field of e-maintenance classified according to required capabilities (i.e. Remote, Collaborative, Predictive maintenance and Knowledge capitalization )and needs intended to be responded ( i.e. Security/reliability of data, Interoperability, Maintenance integration, Collaboration/MAS, Processes formalization and Knowledge management). According to the mentioned report, few platforms such as TELMA and DYNAMITE undertake works on the whole specter of the chosen capabilities with respect to corresponding needs. Interesting case of IMS/D2B ™ is reported to be focusing specifically on predictive maintenance and maintenance integration.

The mosaic of contributions to enhance e-maintenance components with new AI based capabilities within known platforms are then rarely held in a holistic predefined manner. The way intelligence dimension is led bears yet a resemblance to a kind of more or less successful montage upon given ICT infrastructure.

## 3.  Proposal:  Towards  intelligence  based  e-maintenance models

Intelligence is the base of capabilities intended from e-maintenance concept at every business level strategic, tactical or operational levels. Hence, a global view of this ubiquitous dimension enhances mastering its contribution to global and even changing goals designed for an e-maintenance system.

In this work, our intention is to contribute to converging efforts to satisfy real capabilities expected from e-maintenance as an intelligent system within a holistic approach towards an intelligence based framework. This can be understood as a framework where strategy aliened business view is immediately used to design a platform independent intelligence view and then, platform specific and ICT views identify technical possibilities to implementation. Such a framework is intending consensus between globalism in generally Zachman based frameworks and specialism of those focused on limited capabilities and needs. In this work, we focus on the platform independent intelligence view.

In the general Zachman based framework proposed in (Iung et al. 2009), intelligence was mainly mentioned within the organization layer that must support the business process integration. After the mentioned paper, the organization expected for e-maintenance (based on collaborative–cooperative aspects) should be close to the IMS (Intelligent Manufacturing System) paradigm (Hiroyuki, and Yoshikawa. 1995) stating that the system behavior emerges through the dynamics of the interactions of basic maintenance agents within the maintenance environment. Among IMS weaknesses, modeling is mentioned in some literature as in (Thomas, Philippe, and André Thomas. 2011). After these  authors,  that weakness is due to the lack of uniformity in criteria to achieve modeling which makes it very difficult to compare different applications proposed in literature. Our proposal must then deal with such modeling weakness and permit comparison between different proposed e-maintenance solutions.

In (Jacek Jakieła, Bartosz Pomianek .2009), authors argue that agent orientation may be considered as a powerful paradigm for organization modeling and the reference architecture for Management Information Systems, that if properly applied, would lead to firm's overall performance improvement.

A Multi-Agent System is defined by the authors as a set (society) of decentralized software components, that are carrying out tasks collaboratively in order to achieve a goal of the whole society:  those characteristics are among number of similarities between structural and behavioral characteristics of modern organizations and multi-agent systems.

Consequently, we will think the whole e-maintenance system as an intelligent system –i.e. a system with artificial intelligence (Intelligent system – Wikipedia. 2012)- interacting with other constituents of the enterprise as part of its environment to achieve the global enterprise goals. At this point, our work reconsider in a direct and different manner the suggestion of "distributed artificial intelligence environment" when referring to e-maintenance in (Crespo Marquez, Adolfo, and Jatinder N.D. Gupta.2006). Agent metaphor will then be used in this paper to model e-maintenance system as a whole.

Agent definition as well as agent modeling differs from an AI searcher to another according to their own objectives, presuppositions and conceptual frameworks. Sloman, Aaron, and Matthias Scheutz. (2002) introduce a framework for comparing agent architectures: the general CogAff architecture is based on superimposing (1) the distinction between perceptual, central and action components, and (2) a distinction between types of components which evolve at different stages and provide increasingly abstract and flexible processing mechanisms. The reactive components generate goal seeking reactive behavior, whereas the middle layer components enable decision making, planning, and deliberative behavior. The modules of the third layer support monitoring, evaluation and control of the internal process in other layers and levels.

We retain CogAff architecture as the main candidate to stand our proposal due to (1) its founding intention to serve as comparison tool of different solutions, and(2) similarities between its three level hierarchy and the three level of business activities here above mentioned.

Furthermore, according to (Yan et al. 2003), agent roles and role models provide a vocabulary for describing agent systems, with each role describing a position and a set of responsibilities within a certain context or role model. This approach encourages us to think of the problem in terms of roles that need to be played, and the responsibilities associated with each role. Hence, in this section, each component of the CogAff architecture (perception, reactive, deliberative and reflexive layers and action) will be thought as the role intended from that component from a conceptual point of view. Multiple such roles may be assigned to a real agent within particular implementation. This way, the main objective of classification to meet the largest ways of implementations is respected.

The proposed frame, referred to as E-CogAff in the figure here after (fig.1), summarizes our vision to e-maintenance system as an intelligent system called to be more aware of its entire -if not extended- environment. The rest of this section gives details of the frame.

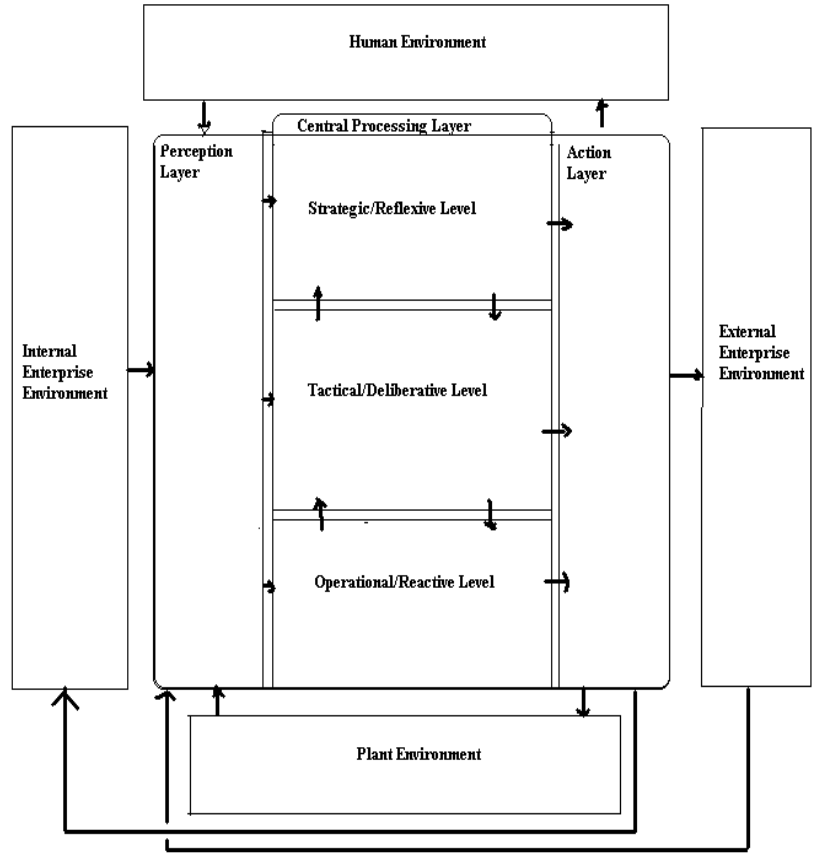

**Fig. 1** E-CogAff : an application of CogAff to e-maintenance system as an intelligent system.

### 3.1. Perception Layer

The role of the perception horizontal layer is responsible of providing central processing layer with information based on data received from different Environment sources. Here, data are to be processed in a hierarchical manner and categorized into different levels of abstraction allowing the central layer to assess the situation of its environment and to make appropriate decisions.

The most direct intelligent system abilities that are candidates to be classified within the perception horizontal layer are:

- Data acquisition and pre-processing to detect and remove discrepancies such as outliers and missing data.
- Data reconciliation
- Data fusion that will help the planning layer assess the situation of the external environment and to make appropriate decisions.

- Database management system

In general, the perception layer must deal with all aspects of data processing and categorizing issues. Data are either manually (e.g. Murphy, Glen D. 2009) or automatically (e.g. Gilabert et al. 2011) collected. Therefore, data quality is of a major concern within the perception subsystem. Several research studies have indicated that most organizations do have data quality problems (Lin et al. 2006).

Accuracy, Completeness, Currency, Consistency and reliability are some of the common dimensions in literature dealing with data quality assessment within information systems as in (Piprani, Baba, and Denise Ernst. 2008). In fact, data quality dimensions characterize properties that are inherent to data. Such dimensions concern data values as well as logical schema or data format to satisfy specific set of semantic rules. Source reliability, defined as the credibility of a source organization with respect to provided data quality values, depends on the cooperative context in which data are exchanged.  In general, security issues are to be tackled within this layer.

## 3.2. Central Processing Layer

The central processing horizontal layer is composed with three hierarchical levels corresponding to strategic, tactical and operational levels of both internal capabilities of the system and different activities of the e-maintenance processes.

### Strategic/Reflective Level

The Strategic/Reflective Layer is responsible of providing the ability to monitor, evaluate, and control other components of the architecture. Assuring required level of awareness of the system environment, this layer is to notice and categorize suspicious situations, and through deliberation or observation over an extended time period develop a strategy to deal with such situations. Furthermore, this level coordinates other modules so as to make the whole system performance more robust and coherent assessing credibility of each module's behavior by monitoring their internal states.

In addition to monitoring, evaluating and control abilities regarding both internal behavior and e-maintenance activities performances , this level is also responsible of assistance in a cooperative manner for maintenance strategic managers. Its responsibility field covers maintenance contribution in an Executive Management Information System as well as interactive search for appropriate strategic goals and plans to maintenance activities within changing internal and external enterprise Environment. Such ability is of great help for maintenance managers dealing with uncertainty- especially within a risk based management approach (Söderholm, Peter, and Ramin Karim. 2011)- enabling a linkage of e-maintenance to the strategic goals of an organization. In such a case, artificial intelligence paradigms in this layer are to address issues concerning some kind of subjective and uncertain aspects of strategic management. Probabilistic methods as well as possibilistic ones such as fuzzy logic based algorithms are among solutions to deal with a certain level of such aspects.

**Tactical/Deliberative Level**

Proactive behavior is achieved in the system in its deliberative layer, which is responsible for governing the system's actions in normal and abnormal circumstances. The choice of the appropriate artificial intelligence paradigm is very crucial to the high performance and real-time requirements of this layer to be endowed with appropriate abilities of planning, reasoning, learning and problem solving: Expert systems, case-based reasoning systems, neural networks are among solutions for such a requirement with different strengths and weaknesses to be taken into account in a given context.

Such artificial intelligence requirements are to address different issues including exceptional processing regarding time or resources required by certain maintenance policies: reliability centered maintenance, opportunistic maintenance, health/care management, knowledge management and asset life cycle management. In general, in addition to the supervision of the system in coordination with the reflective layer, this layer is also responsible of interactive decision support towards tactical maintenance managers.

**Operational/Reactive Level**

The reactive layer provides direct responses to events that occur in the environment. Inspection and condition based maintenance is the major element of this layer. Many artificial intelligence paradigms can coexist in this layer with appropriate mechanism for collaborative problem-solving concept. This layer is also to support maintenance staff in their daily activities providing them with the right information at the right time and the right place, asking for required data and alerting in case of errors or lack of provided data among other interactive functionalities.

### 3.3. Action Layer

The action layer concretizes plans that are decided in the central layer within a hierarchically organized manner. Among responsibilities of this layer are Maintenance Planning and scheduling, different presentation tasks towards human Environment, updating adequate internal and external Environment and eventually acting on the plant Environment.

### 3.4. Environment

The intelligent system is intended to act on its perceived environment in a way to achieve its maintenance goals and thereby enterprise common goals. We decompose such an environment into four categories as follows:

**Human Environment**

This includes all human agents that are to interact with the system in relation to maintenance. Be it management, executive or functional and technical staff or human experts, contractor's team members, constructor's after-sale responsible and so on. Some interactive activities are listed within the three levels of the central layer. Interactivity can be understood as simple information exchange as well as the way the system enables supporting individual maintenance team members in completion of their own tasks or the team as a whole.

**Plant Environment**

This environment includes mainly:
- Sensors regarding physical assets condition (vibration, oil analysis…) as well as operational conditions (temperature, voltage,...)
- Actuators in case of integrated control (emergency shutdown,…) or self-maintainable assets.

**Enterprise Environment**

In the enterprise environment, a distinction is made between internal and external environment to allow useful differentiation in the system perception regarding those environments. External environment needs more prudence such as evaluation of the credibility of a source organization.

*Internal Enterprise Environment*
The internal enterprise environment includes every system fully accessible to the maintenance system in a manner to allow it adjust its goals and its way to achieve them, as well as contributing in the achievement of the common goal of the enterprise.

The main department in daily direct relation with maintenance are typically: Human resources (availability, skills, training, salaries…), Production (operational plan…), MRO Inventory and Purchasing (Maintenance Repair Operations requirement), Finances (costs….)

Regardless to their geographical position or the level of their intelligence, internal systems include Management and Executive Information Systems, Asset Management Systems, Enterprise Resource Planning and CMMS

*External Enterprise Environment*
External environment include every system not owned by the enterprise offering services to the maintenance system or cooperating occasionally or permanently to achieve predefined goals.

Following systems are examples: Inter Enterprise databases such as reliability databases of a common industrial sector, external knowledge and experts systems, external catalogues and machines manufacturer's services that do not introduce human agents. This includes the case of exchange of reliability data for and from maintenance system especially when redesign is needed.

## 4.   Case study: DINAWEB

In this section, we'll use our proposed frame to demonstrate its applicability in studying the case of Dynaweb platform: an ICT architecture related to software web services and communication architecture that support the e-maintenance concept within an integrated European project Dynamite (Dynamic Decisions in Maintenance). With its 16 partners, Dynamite Project is a real leader promoting e-maintenance concept either in (1)developing condition monitoring sensors and smart tags, (2) introducing mobile & wireless devices and technologies to support maintenance and diagnosis and prognosis based on WebServices or  (3)developing training and economical studies related to maintenance strategy.

This section is mainly based on information given in (Jantunen et al. 2009) as reported within the 4th World Congress on Engineering Asset Management, 28-30 September 2009, Athens, Greece. In that conference, important results of Dynamite were discussed with reference to project deliverables here after referred to as DWC followed by the number of some of them.

The software architecture of DynaWeb is based on software components offering intelligent distributed web services and was demonstrated in architectures with logically structured decision layers based on the open systems architecture for CBM definition, from condition monitoring to decision support, and provides automated extraction of results(Irigaray et al. 2009). As   explained in the main source of this section, different software modules are able to communicate with each other in order to perform a specific task. Indeed, in this context of Dynaweb, information processing is understood as a distributed and collaborative system, where there are different levels of entities that can undertake intelligent tasks. Hence, this case presents features of e-maintenance as an intelligence system as required by our proposed frame.

The perception layer is materialized within Dynaweb by its MIMOSA database which follows the ISO 13374 standard: Machine Condition Assessment Data Processing Information Flow Blocks.

The central processing layer is based on corresponding layers of OSA-CBM standard  that are condition monitoring , health assessment, Prognostics and Decision support . Within this central we can recognize following constituents:

The operational/reactive level of our frame that corresponds to DWC19   is related to state detection: Measurements received from sensors and their signal processing software are compared to expected values, and alerts are generated in case of anomaly detection.

The tactical/deliberative level is well developed within that project. In case of an anomaly detected by condition monitoring modules DWC20   generates health records. Faults are then identified based on health history, operational status and maintenance tasks history.  Primary focus of the prognostic module DWC21 is to calculate the future health of an asset and report remaining useful life, taking into account results from DWC 19 and DWC20.

The strategic/reflexive level is yet focusing on the only maintenance costs aspect instead of a more global vision based on what maintenance provides and what are its global purposes and benefits. DWC24 refers to a set of modules developed within the Dynamite project for enabling economical analysis related to maintenance.

Action layer in Dynaweb is mainly presentations, messages and generally information submitted to the destination of different environment constituents. In addition to these functionalities, special attention was given to maintenance scheduling in DWC13 and DWC23.

Plant Environment is by far the most important field concentrating efforts within dynamite project since its starting in 2005. A number of new sensors were developed in this project including vibration (DWC12) and on-line oil analysis sensors. In DWC3 , a multi sensors approach for monitoring vibration, pressure and temperature was concretized based on MEMS technology. Also, a wireless sensor network system, including ZigBee sensing nodes and Zigbee to Wifi gateways, was developed within DWC15 and DWC18. For on-line monitoring of lubricating oil five different types of sensors were developed within DWC4 to DWC8. Most of the developed sensors are based on optical methods. In addition, RFID tags was studied within Dynamite project for the identification of machines to be maintained (DWC1, DWC10, DWC2 and DWC11) and the report mentioned that more research are still needed for commercial use.

In relation to the Human Environment, PDAs play a central role in the Dynamite project. The report explains that the PDA are used for accessing measurements data, studying the measurement results, communicating with the diagnosis and prognosis modules, handling of work orders, studying instructions for maintenance work and so on. Furthermore, in DWC9 it is assumed that it must be possible to use the PDA to support maintenance even if wireless connection is not available.

Integrating e-maintenance system with its Enterprise Environment is considered within Dynamite project as a key to the success of current e-Maintenance solutions. Referring to the mentioned report, information integration has been pursued by adopting interoperable XML based data exchange formats, operational level data exchange protocols and maintenance-oriented data format definitions. Finally, networking integration has been also pursued through the combined use of wired and wireless communications. However, issues related to performance degradation due to wireless networking are mentioned in the report. Furthermore, CMMS and ERP as essential constituents of e-maintenance system enterprise environment are reported still not fully addressed. In deed, many CMMS and ERP solutions do not necessarily support required data interoperable formats.

Finally, it's useful to end this case study with some results of industrial demonstrations related to DynaWeb solution. The aim of such industrial demonstrations was to integrate the technological and information technology strands in a business sense and to verify the effectiveness of the complete DynaWeb solution as reported in (Mascolo et al. 2010):

The overall result is reported to be globally positive, with technical and economical feasibility proven. However, in an interesting case where the feasibility of integrating the DynaWeb components to form the e-maintenance architecture has been tested on the TELMA Platform, integration issues were still reported. The TELMA was developed mainly for supporting e-maintenance purposes, from both educational and research points of view. According to the same report, tests of feasibility and architecture aspect were held into (1)Internal test, to test the function of each ICT component of the DynaWeb platform and validate its ability to tackle real data, and (2) Integration test, to integrate all components and validate the ability of the Dynaweb platform to tackle an e-maintenance problem from sensor detection, to high level decision and scheduling.

Concerning the internal tests, the results are reported to be positive while the results of the integration tests are less positive. According to the same report, this is mainly due to (1) MIMOSA compliancy was not ensured for all components, and (2) Missing functionalities according to use case requirements. Adjustments were required from the developer's or the end-user's point of view to assign the components with the right capacities needed for the final implementation on site. This last issue gives an idea of the further long way ahead to a real integration of e-maintenance as an intelligent system into human teams.

## 5. Discussion and future work

The ISO 13374 is a condition monitoring architecture that consists of 8 functional blocks:

The first two blocks are considered within the perception layer of our proposed frame, namely:

- Data acquisition: Produces digital records with values and descriptive meta-information like time stamps and quality for the sensor/transducer output.

- Data manipulation: Performs any kind of processing (e.g., signal processing) necessary in order for the sensor values to be meaningful with respect to state and health assessment.

Follows, a reactive layer component:

- State detection: Detects whether the asset is in a normal or abnormal state.

The next three blocks implement the diagnosis and remediation intelligence of the system. These blocks are deliberative component:

- Health assessment: Rates the current state as asset health and diagnoses any faults.

- Prognostic assessment: Predicts the remaining lifetime until the next significant state change.

- Advisory generation: Provides the user with recommendations on maintenance actions or changes of operational settings to optimize asset performance in the given state.

The last blocks are action responsibilities towards some of the environment composers:

- Information presentation: Presents the output of any of the previous stages to the user in easily understandable, textual, and graphical form. This block is among responsibilities of the action layer towards human environment.

- External systems: Any external system the asset management application is connected to such as archiving, configuration/engineering and history. This block belongs to responsibilities of the action layer towards internal enterprise environment. External enterprise environment is not necessary excluded. (Naedele, M., P. Sager, and C. Frei.2004)

As it is the case of Dynaweb here above studied in detail, this standard is at the base of several condition based maintenance solutions in the e-maintenance field. Thus, the first utility of the proposed frame is to provide some kind of classification - as inherited from the founding intention of its CogAff base - of at least architectures based on the mentioned standard.

Focusing on condition monitoring, the standard has evident influence on the intelligence dimension of the corresponding solutions. Indeed, this dimension hardly goes beyond the deliberative layer neither deals with actions toward human environment within a real integration of the intelligent system into human teams. In fact, the predominance of the definition of e-maintenance as a part of e-manufacturing and e-business pushed such architectures to focus on more automation and less human intervention. However, to improve the classification ability of the proposed frame, further work is needed to enhance exhaustiveness regarding responsibilities assigned to each role in the proposed frame.

The second usefulness of the frame is to map existing intelligence abilities within an e-maintenance system in a given plant or platform and have an idea about intelligence requirements not satisfied. The projected intelligence level deduced from a requirement study can as well be mapped within that frame. This utility at the conceptual level of an e-maintenance project is at the core of our future work towards an intelligence based framework for e-maintenance.Related to this conceptual helpfulness investigation, we note (Sayda, Atalla F., 2008): one of the rare direct applications of the CogAff approach is described in that paper with an intelligent industrial production control system (ICAM). The proposed architecture of the conceptual system consists of four information processing layers and three vertical subsystems, namely, perception, central processing, and action. The lowest horizontal layer above the distributed control system (DCS) contains semi-autonomous agents that represent different levels of data abstraction and information processing mechanisms of the system. The middle two layers interact with the external environment via the DCS by acquiring perceptual inputs and generating actions. The perceptual and action subsystems are divided into several layers of abstraction to function effectively. While ICAM case comfort the feasibility of using the proposed E-CogAff frame in the conceptual stage of an e-maintenance project, in our future work we will attempt to tackle with the two main issues here above underlined: (1) more integration of the e-maintenance system at the strategic level of the enterprise and (2) more integration of e-maintenance as an intelligent system within human teams.

## 6. Conclusion

This paper underlined the importance of intelligence as a fundamental dimension of an e-maintenance system. First, a review of the three level of maintenance management aligned with European standards is given. Then, e-maintenance is presented as an ICT supported information system. With a review of literature and known e-maintenance platforms and framework, we discussed the importance explicitly given to the intelligence of e-maintenance systems. According to our findings, this dimension suffers from the lack of frameworks explicitly dealing with the intelligence of e-maintenance systems in a holistic approach.

As our contribution towards intelligence based conceptual framework for e-maintenance, we proposed a frame sustaining the eligibility of CogAff architecture as candidate for such a task. Our proposed frame aims to serve as (1) a classification means of existing e-maintenance solutions and (2) a platform independent conceptual tool within an e-maintenance project. To illustrate classification ability of the proposed frame, an application to the case of an e-maintenance leader –Dynaweb- was detailed. Finally, we discussed opportunities and limitations of our proposal. A brief application to ISO 13374 based e-maintenance solutions allowed us to highlight issues regarding (1) real integration of e-maintenance as an intelligent system into human teams, and (2) more mature implication at the strategic level of management. To serve as an accurate tool for e-maintenance solutions classification, our proposal needs further work to enhance its exhaustiveness. At the conceptual level, in our future work within an e-maintenance project we will attempt to tackle with the two main integration issues here above mentioned (ie- at the strategic level and within human teams).